# Microburst Nowcasting Applications of GOES

KENNETH L. PRYOR

Center for Satellite Applications and Research (NOAA/NESDIS), Camp Springs, MD

## ABSTRACT

Recent testing and validation have found that the Geostationary Operational Environmental Satellite (GOES) microburst products are effective in the assessment and short-term forecasting of downburst potential and associated wind gust magnitude. Two products, the GOES sounder Microburst Windspeed Potential Index (MWPI) and a new bi-spectral GOES imager brightness temperature difference (BTD) product have demonstrated capability in downburst potential assessment. In addition, a comparison study between the GOES-R Convective Overshooting Top (OT) Detection and MWPI algorithms has been completed for cases that occurred during the 2007 to 2009 convective seasons over the southern Great Plains. Favorable results of the comparison study include a statistically significant negative correlation between the OT minimum temperature and MWPI values and associated measured downburst wind gust magnitude. The negative functional relationship between the OT parameters and wind gust speed highlights the importance of updraft strength, realized by large CAPE, in the generation of heavy precipitation and subsequent intense convective downdraft generation. This paper provides an updated assessment of the GOES MWPI and GOES BTD algorithms, presents case studies demonstrating effective operational use of the microburst products, and presents results of a cross comparison study of the GOES-R overshooting top (OT) detection algorithm over the United States Great Plains region.

________________

## 1. Introduction

Recent testing and validation have found that the Geostationary Operational Environmental Satellite (GOES) microburst products are effective in the assessment and short-term forecasting of downburst potential and associated wind gust magnitude. Two products, the GOES sounder Microburst Windspeed Potential Index (MWPI) and a new two-channel GOES imager brightness temperature difference (BTD) product have demonstrated capability in downburst potential assessment (Pryor 2008; Pryor 2010). The GOES sounder MWPI algorithm is a predictive linear model developed in the manner exemplified in Caracena and Flueck (1988):

$$\text{MWPI} \equiv \{(\text{CAPE}/100)\} + \{\Gamma + (T-T_d)_{850} - (T-T_d)_{670}\} \quad (1)$$

*Corresponding author address*: Kenneth L. Pryor, Center For Satellite Applications and Research, 5200 Auth Road, Camp Springs, MD 20746
E-mail: Ken.Pryor@noaa.gov

where CAPE represents Convective Available Potential Energy, $\Gamma$ is the lapse rate in degrees Celsius (°C) per kilometer from the 850 to 670-mb level, and the quantity $(T-T_d)$ is the dewpoint depression (°C). In addition, it has been found recently that the BTD between GOES infrared band 3 (water vapor, 6.5μm) and band 4 (thermal infrared, 11μm) can highlight regions where severe outflow wind generation (i.e. downbursts, microbursts) is likely due to the channeling of dry mid-tropospheric air into the precipitation core of a deep, moist convective storm. Rabin et al. (2010) noted that observations have shown that BTD > 0 can occur when water vapor exists above cloud tops in a stably stratified lower stratosphere and thus, BTD > 0 has been used a measure for intensity of overshooting convection. A new feature presented in this paper readily apparent in BTD imagery is a "dry-air notch" that signifies the channeling of dry air into the rear flank of a convective storm.

In addition, a comparison study between the GOES Convective Overshooting Top (OT) Detection and MWPI algorithms has been completed for cases that occurred during the 2007 to 2009 convective seasons over the southern Great Plains. The OT detection algorithm (Bedka et al. 2010) is a pattern recognition-based technique that employs brightness temperature (BT) data from the GOES thermal infrared channel. Output OT detection algorithm parameters include cloud top minimum BT and a BT difference between the overshooting top and surrounding convective anvil cloud. Favorable results of the comparison study include a statistically significant negative correlation between the OT minimum temperature and MWPI values and associated measured downburst wind gust magnitude. The negative functional relationship between the OT parameters and wind gust speed highlights the importance of updraft strength, realized by large CAPE, in the generation of heavy precipitation and subsequent intense convective downdraft generation. This paper will provide an updated assessment of the GOES

MWPI and GOES BTD algorithms, presents case studies demonstrating effective operational use of the microburst products, and presents results of the intercomparison study of the GOES-R overshooting top (OT) detection algorithm over the United States Great Plains region.

## 2. Data collection and methodology

The main objective of the validation effort is to qualitatively and quantitatively assess and intercompare the performance of the MWPI, BTD, and OT algorithms by employing classical statistical analysis of real-time data. Algorithm output data was collected for downburst events that occurred during the warm season (especially between 1 June and 30 September) and was validated against surface observations of convective wind gusts as recorded by surface observing stations in mesoscale networks in Oklahoma, western Texas, and the Atlantic coastal region. Site characteristics, data quality assurance, and wind sensor calibration are thoroughly documented in Brock et al. (1995) and Schroeder et al. (2005). Wakimoto (1985) and Atkins and Wakimoto (1991) discussed the effectiveness of using mesonetwork surface observations and radar reflectivity data in the verification of the occurrence of downbursts.

As illustrated in the flowchart in Fig. 1, MWPI product images are generated by Man computer Interactive Data Access System (McIDAS)-X by a program that reads and processes GOES sounder data, calculates and collates microburst risk values, and overlays risk values on GOES imagery. For selected downburst events, the MWPI product was generated using the Graphyte Toolkit. The MWPI algorithm, as implemented in the Graphyte Toolkit, reads and processes GOES sounder profile data in binary format available on the GOES sounding profile web page (http://www.star.nesdis.noaa.gov/smcd/opdb/goes/soundings/html/sndbinary23L.html). The MWPI is then calculated for each retrieval location and plotted as a colored marker on a user-defined map.

For the BTD product, image data consisted of derived brightness temperatures from GOES-East (GOES-12, 2009 and before; GOES-13, 2010 and after) 4-km resolution water vapor (band 3) and thermal infrared (band 4), obtained from the Comprehensive Large Array-data Stewardship System (CLASS, http://www.class.ncdc.noaa.gov/). Microburst algorithm output was visualized by McIDAS-V software (Available online at http://www.ssec.wisc.edu/mcidas/software/v/). A contrast stretch and built-in color enhancement ("Pressure") were applied to the output images to highlight patterns of interest including overshooting tops and dry-air notches. Next Generation Radar (NEXRAD) and Terminal Doppler Weather Radar (TDWR) base reflectivity imagery from National Climatic Data Center (NCDC) were utilized to verify that observed wind gusts are associated with high-reflectivity downbursts and not associated with other types of convective wind phenomena (i.e. gust fronts). Particular radar reflectivity signatures, such as the rear-inflow notch (RIN)(Przybylinski 1995) and the spearhead echo (Fujita and Byers 1977), are effective indicators of the occurrence of downbursts.

In order to assess the predictive value of the algorithm output, the closest representative index values were obtained for retrieval times one to three hours prior to the observed surface wind gusts. A technique devised by Wakimoto (1985) to visually inspect wind speed observations over the time intervals encompassing candidate downburst events was implemented to exclude gust front events from the validation data set. Algorithm effectiveness was assessed by computing the correlation between MWPI values, OT temperature and observed surface wind gust velocities. Statistical significance testing was conducted to determine the confidence level of the correlation between observed downburst wind gust magnitude and microburst risk values.

## 3. Analysis

*a.    August 2009 Oklahoma Downbursts*

During the afternoon of 10 August 2009, strong convective storms developed along a cold front that extended from eastern Kansas to the Oklahoma Panhandle.  The storms tracked eastward and produced several strong downbursts over northwestern and north-central Oklahoma during the late afternoon and evening.  Table 1 features a listing of five significant high wind measurements by Oklahoma Mesonet stations during this downburst event.  The last and highest downburst wind gust of the event, 33 ms$^{-1}$ (64 kt), was recorded at Freedom station at 2345 UTC.  It was found that the GOES MWPI, BTD and OT detection algorithms were effective in indicating the intensity of convective storm activity and resulting downburst wind gust magnitude.  A correlation of 0.83 between MWPI values and measured wind gusts and a correlation of -0.88 between OT minimum BT and measured wind gusts exemplified the strong statistical relationship between these parameters and downburst wind gust magnitude.

At 2100 UTC, Fig. 2, the MWPI product image displayed elevated values widespread over Kansas and Oklahoma with high values (red) that indicate wind gust potential greater than 50 knots extending from northeastern to southwestern Oklahoma, well ahead of the developing convective storm activity.  Along and immediately ahead of the cold front, moderate (yellow) values indicated wind gust potential of 18 to 25 ms$^{-1}$ (35 to 49 kt).  Downburst wind gusts recorded by mesonet stations between 2115 and 2305 UTC ranged from 20 to 23 ms$^{-1}$ (40 to 45 kt), consistent with gust potential as shown in the 2100 UTC MWPI product.  By 2200 UTC, MWPI values increased significantly downstream of a new area of developing storm activity along the cold front.  An intrusion of dry mid-tropospheric air was becoming apparent on the western flank of the developing convective storm complex over the eastern Oklahoma Panhandle and extreme southern Kansas.  Figure 2c, the Graphyte visualization of the 2200 UTC MWPI, shows moderate values (36 to 42, green markers) in the vicinity of Dodge City, Kansas (DDC), the location of the 0000 UTC radiosonde observation (RAOB) displayed in Fig. 3.  Figure 2c is also effective in highlighting that the Freedom downburst occurred in proximity to a local maximum in MWPI (near 60, red marker) just north of the Kansas-Oklahoma border.  Dry air notches had become especially well-defined on the northern and western flanks of the convective storm as shown in the 2332 UTC BTD image in Fig. 3.

Note that in Fig. 3, the overshooting top was co-located with a maximum in radar reflectivity and located 13 km northwest of the Freedom mesonet station, where a downburst wind gust of 33 ms$^{-1}$ (64 kt) was recorded between 2340 and 2345 UTC.  The cold overshooting top BT value of 196°K was associated with the severe downburst, highlighting the importance of convective storm updraft strength in the generation of severe weather, especially high winds. The closest representative MWPI value of 52 corresponded to wind gust potential of 25 to 33 ms$^{-1}$ (50 to 64 kt).  A dry air notch, apparent in the BTD image as a light-to-dark blue shaded indentation, illustrated the role and importance of the entrainment of dry, mid-tropospheric air into the rear flank of the convective storm and the subsequent generation of intense downdrafts. The favorable environment for downbursts was well illustrated in the 0000 UTC (11 August) radiosonde observation from Dodge City, Kansas.  Large CAPE and a large temperature lapse rate below the 550-mb level definitely were instrumental in fostering a strong storm updraft that was realized as a heavy precipitation core and overshooting top observed at 2332 UTC.  In addition, the sounding indicated the presence of two prominent dry-air layers centered near the 400-mb and 700-mb levels.  Strong westerly winds of 20 to 25 ms$^{-1}$ (40 to 50 kt) were measured in the upper dry-air layer that likely injected this dry air into the storm precipitation core that enhanced downdraft acceleration and subsequent severe winds observed on the surface by the Freedom mesonet station. All of these favorable conditions were reflected in the 2332 UTC BTD image that displayed concurrent high storm radar reflectivity, an overshooting top, and a prominent dry-air notch on the western flank of the storm.

*b.    May 2011 Hampton Roads Downbursts*

During the afternoon of 24 May 2011, a multicellular convective storm developed over the southern piedmont of Virginia and tracked rapidly eastward toward the lower Chesapeake Bay. Between 2000 and 2100 UTC, as the convective storm passed over the Hampton Roads, one of the busiest waterways in the continental U.S., numerous severe wind gusts were recorded by WeatherFlow (WF) and Physical Oceanographic Real-Time System (PORTS) stations. After inspecting satellite and radar imagery for this event, it was apparent that these severe wind observations were associated with downburst activity. GOES MWPI imagery in Fig. 4 indicated a general increase in wind gust potential over the Hampton Roads area during the afternoon hours, between 1900 and 2000 UTC. The increase in both convective and downdraft instability was reflected in the Norfolk, Virginia GOES sounding profile in Fig. 5 as a marked increase in CAPE and an elevation and increasing amplitude of the mid-tropospheric dry-air layer. By 2000 UTC, the MWPI product indicated the highest wind gust potential, up to 33 ms$^{-1}$ (64 kt), over Hampton Roads, where wind gusts of 29 to 35 ms$^{-1}$ (57 to 67 kt) were recorded by WeatherFlow and PORTS stations during the following hour. Table 2 lists the most significant wind observations associated with severe thunderstorm event. Visible imagery emphasized the multicellular structure of the storm with overshooting tops identifying the most intense convective cells that were capable of producing severe downbursts. Figure 6 illustrates the observing network over the Hampton Roads area that revealed the divergent nature of the downburst winds as the storm was tracking overhead. Figure 7a, the BTD product image at 2025 UTC, with overlying radar reflectivity, revealed favorable conditions for severe downbursts with prominent dry-air notches on the southwestern and northwestern flanks of the storm pointing toward the convective precipitation core. This signifies that the entrained mid-tropospheric dry air was interacting with the storm precipitation core to result in evaporational cooling, negative buoyancy generation, and subsequent acceleration of storm downdrafts. Similar to the August 2009 case, GOES sounding profiles reflected elevated MWPI values by displaying the presence of large CAPE, a dry sub-cloud layer, and a steep temperature lapse rate below the 700-mb level. By 2040 UTC, as shown in Fig. 7b, dry-air notches had become more pronounced in BTD imagery while prominent spearhead echoes were apparent in overlying radar imagery. Near this time, significant severe downburst winds were recorded by PORTS and WeatherFlow stations on the Chesapeake Bay Bridge-Tunnel between Virginia Beach and Cape Charles. As the storm moved eastward over the Virginia Beach oceanfront, weaker downburst winds were recorded where wind gust potential of 18 to 25 ms$^{-1}$ (35 to 49 kt) was indicated at 2000 UTC. Graphyte visualizations were not shown for this case due to the lack of data plots over the Hampton Roads area.

## 4. Discussion

Validation results for the 2007 to 2010 convective seasons have been completed for the MWPI product. GOES sounder-derived MWPI values have been compared to mesonet observations of downburst winds over Oklahoma and Texas for 208 events between June 2007 and September 2010. The correlation between MWPI values and measured wind gusts was determined to be 0.62 and was found to be statistically significant near the 100% confidence level, indicating that the correlation represents a physical relationship between MWPI values and downburst magnitude and is not an artifact of the sampling process. Figure 8 shows a scatterplot of MWPI values versus observed downburst wind gust speed as recorded by mesonet stations in Oklahoma and Texas. The MWPI scatterplot identifies two clusters of values: MWPI values less than 50 that correspond to observed wind gusts between 18 to 26 ms$^{-1}$ (35 to 50 kt), and MWPI values greater than 50 that correspond to observed wind gusts greater than 26 ms$^{-1}$ (50 kt). The scatterplot illustrates the effectiveness of the MWPI product in distinguishing between severe and non-severe convective wind gust potential.

The comparison study between the GOES-R Convective Overshooting Top (OT) Detection and MWPI algorithms has also produced favorable results, shown in Fig. 9, that include a statistically significant negative correlation between the OT minimum temperature and MWPI values (r=-0.47) and OT temperature and measured wind gust magnitude (r=-0.39) for 47 cases that occurred between 2007 and 2009. The negative functional relationship between the OT parameters and wind gust speed highlights the importance of updraft strength, realized by large CAPE, in the generation of heavy precipitation and subsequent intense convective downdraft generation. These results are consistent with the findings of Dworak et al. (2011) that show a near- linear relationship between overshooting top BT and magnitude and severe wind frequency (Fig. 9a, b).

## 5. Conclusions

As documented in Pryor (2008, 2010), and proven by statistical analysis, the GOES sounder MWPI product has demonstrated capability in the assessment of wind gust potential over the southern Great Plains and Mid-Atlantic coast regions. Case studies and statistical analysis for downburst events that occurred during the 2007 to 2010 convective seasons demonstrated the effectiveness of the GOES microburst products. However, as noted by

Caracena and Flueck (1988), the majority of microburst days during JAWS were characterized by environments intermediate between the dry and wet extremes (i.e. hybrid).  As noted in Pryor (2008), the MWPI product is especially useful in the inference of the presence of intermediate or "hybrid" microburst environments, especially over the Great Plains region.  Further validation over the Atlantic coast region should strengthen the functional relationship between MWPI values and downburst wind gust magnitude.

The dry-air notch identified in both case studies presented above likely represents drier (lower relative humidity) air that is entrained into the rear of convective storms and interacts with their precipitation cores, subsequently providing the energy for intense downdrafts and resulting downburst winds. Comparison of BTD product imagery to corresponding radar imagery revealed a physical relationship between the dry-air notch and the spearhead echo.  Entrainment of drier mid-tropospheric air into the precipitation core of the convective storm typically results in evaporation of precipitation, the subsequent cooling and generation of negative buoyancy (sinking air), and resultant acceleration of a downdraft. When the intense localized downdraft reaches the surface, air flows outward as a downburst. Ellrod (1989) noted the importance of low mid-tropospheric (500-mb) relative humidity air in the generation of the severe Dallas-Fort Worth, Texas microburst in August 1985.  Thus, the band 3-4 BTD product can serve as an effective supplement to the GOES sounder MWPI product.  Further validation of the imager microburst product and quantitative statistical analysis to assess product performance will serve as future work in the development and evolution of the GOES microburst products.

*Acknowledgements*.  The author thanks the Oklahoma Mesonet, and Jay Titlow (WeatherFlow) for the surface weather observation data used in this research effort.  The author thanks Michael Grossberg and Paul Alabi (NOAA/CREST, CCNY) for their implementation of the MWPI program into the Graphyte Toolkit and their assistance in generating the MWPI product images. The author also thanks Jaime Daniels (NESDIS) for providing GOES sounding retrievals displayed in this paper.

# TABLES AND FIGURES

Table 1.  Measured wind gusts compared to MWPI values and OT properties for the 10 August 2009 Oklahoma downburst event.  All times are in UTC.

| Time | Gust Speed ms$^{-1}$ (kt) | Location | MWPI | OT Time | OT Dist (km) | OT Min (°K) | OT Mag (°K) |
|---|---|---|---|---|---|---|---|
| 2115 | 20.6 (40) | Copan | 31 | 2115 | 9 | 206.5 | -9.2 |
| 2125 | 22.1 (43) | Lahoma | 44 | 2115 | 9 | 210.1 | -7 |
| 2230 | 21.1 (41) | Slapout | 33 | 2215 | 12 | 211.6 | -8.7 |
| 2305 | 23.1 (45) | Buffalo | 44 | 2302 | 13 | 203.5 | -8 |
| 2345 | 32.9 (64) | Freedom | 52 | 2332 | 13 | 196.8 | -11.2 |

Table 2.  Measured wind gusts (knots) for the 24 May 2011 Hampton Roads downburst event.  Time is in UTC.  WeatherFlow stations are identified by "WF".

| Time | Gust Speed ms$^{-1}$ (kt) | Location |
|---|---|---|
| 2015 | 27.3 (53) | Poquoson (WF) |
| 2017 | 29.3 (57) | Monitor-Merrimack Memorial Bridge Tunnel (WF) |
| 2018 | 32.4 (63) | Willoughby Degaussing Station (PORTS) |
| 2020 | 30.4 (59) | Hampton Flats (WF) |
| 2036 | 34.5 (67) | 1$^{st}$ Island (PORTS) |
| 2040 | 31.9 (62) | 3$^{rd}$ Island (WF) |

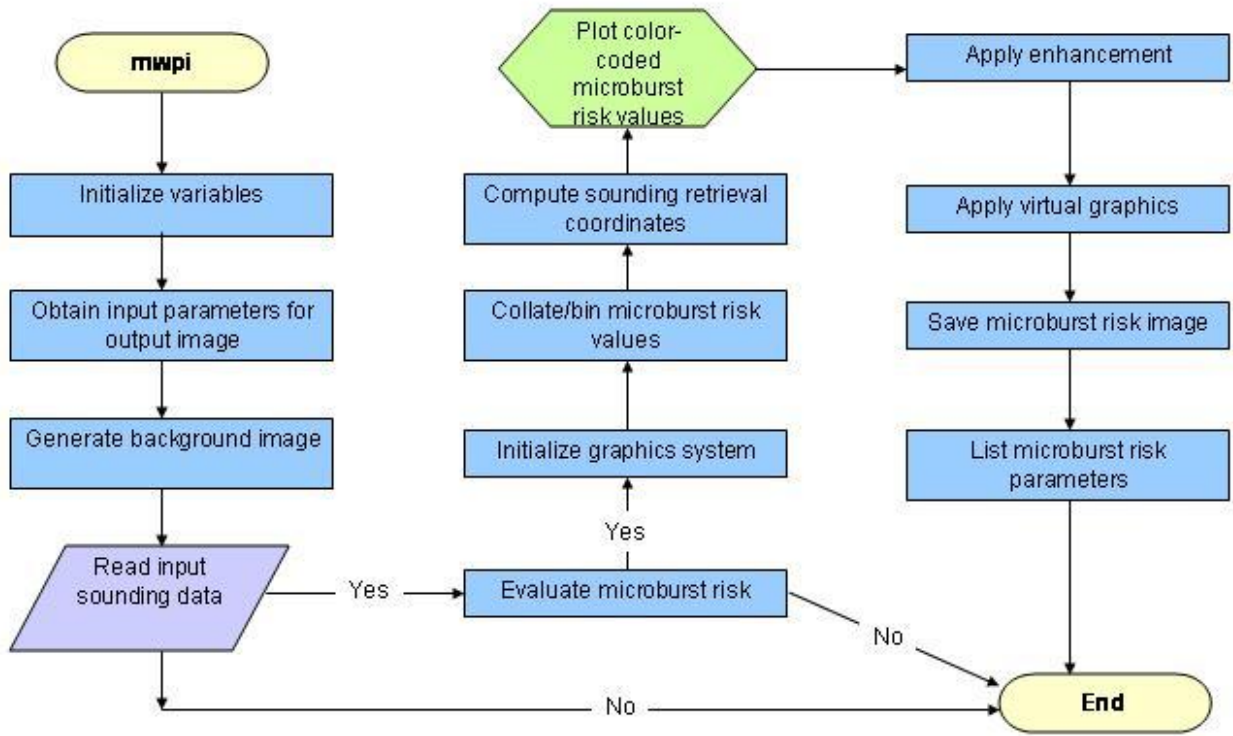

**Figure 1**. Flowchart illustrating the operation of the MWPI program.

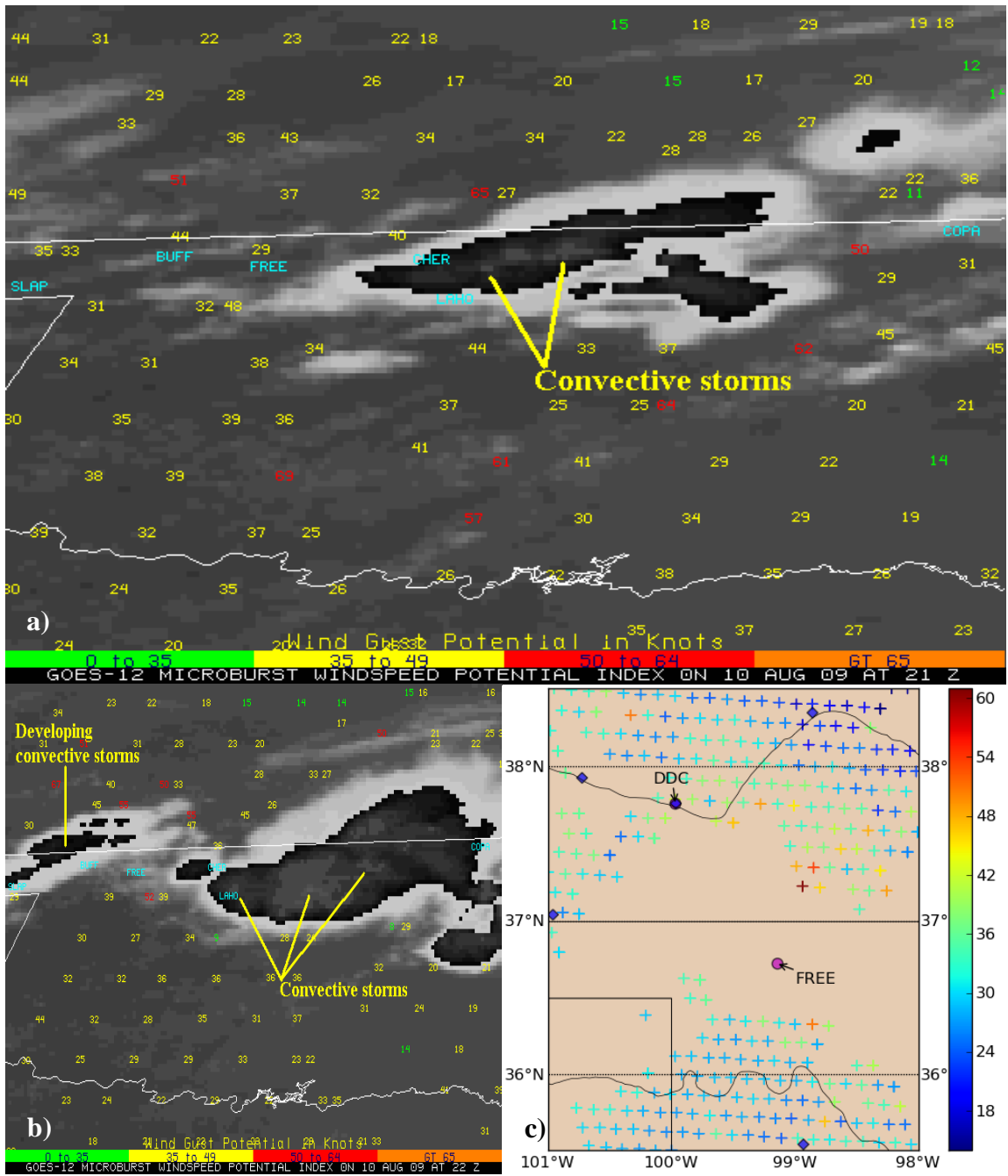

**Figure 2**. Geostationary Operational Environmental Satellite (GOES) Microburst Windspeed Potential Index (MWPI) overlying infrared imagery on 10 August 2009 at a) 2100 UTC, b) 2200 UTC (McIDAS version), and c) 2200 UTC (Graphyte version). Four-letter identifiers of Oklahoma Mesonet stations listed in Table 1 are plotted over the image.

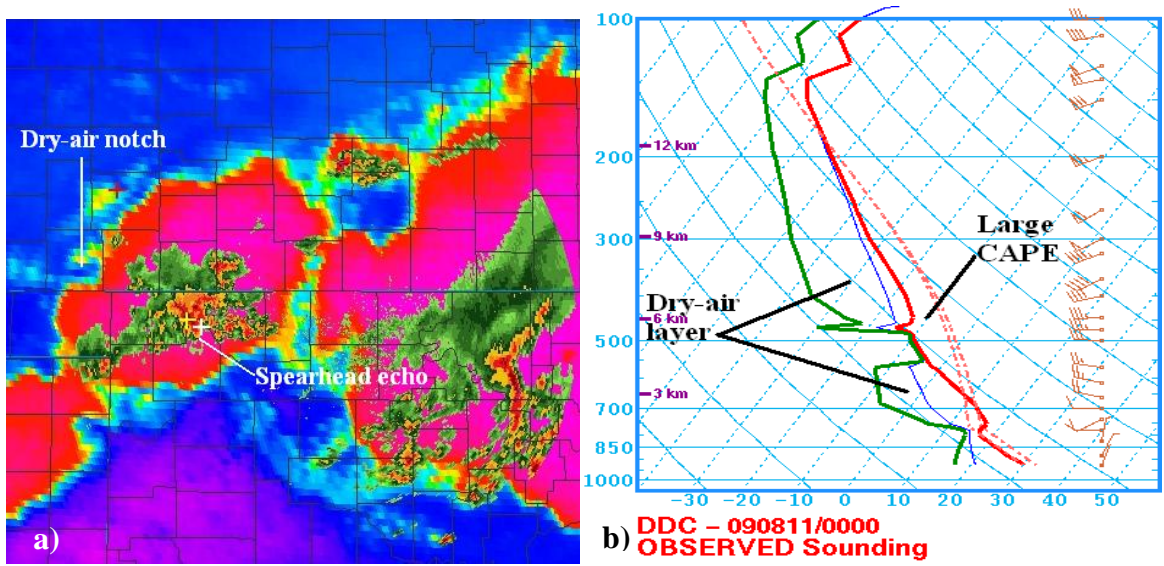

**Figure 3**. a) GOES imager channel 3-4 BTD product at 2332 UTC 10 August 2009, with overlying radar reflectivity from Vance AFB NEXRAD; b) Radiosonde observation (RAOB) over Dodge City, Kansas at 0000 UTC 11 August 2009. The location of downburst occurrence at Freedom, Oklahoma Mesonet station is marked with a white cross. The adjacent overshooting top, as indicated by the Bedka algorithm, is marked with a yellow cross. The location of the Dodge City RAOB is marked with a red cross.

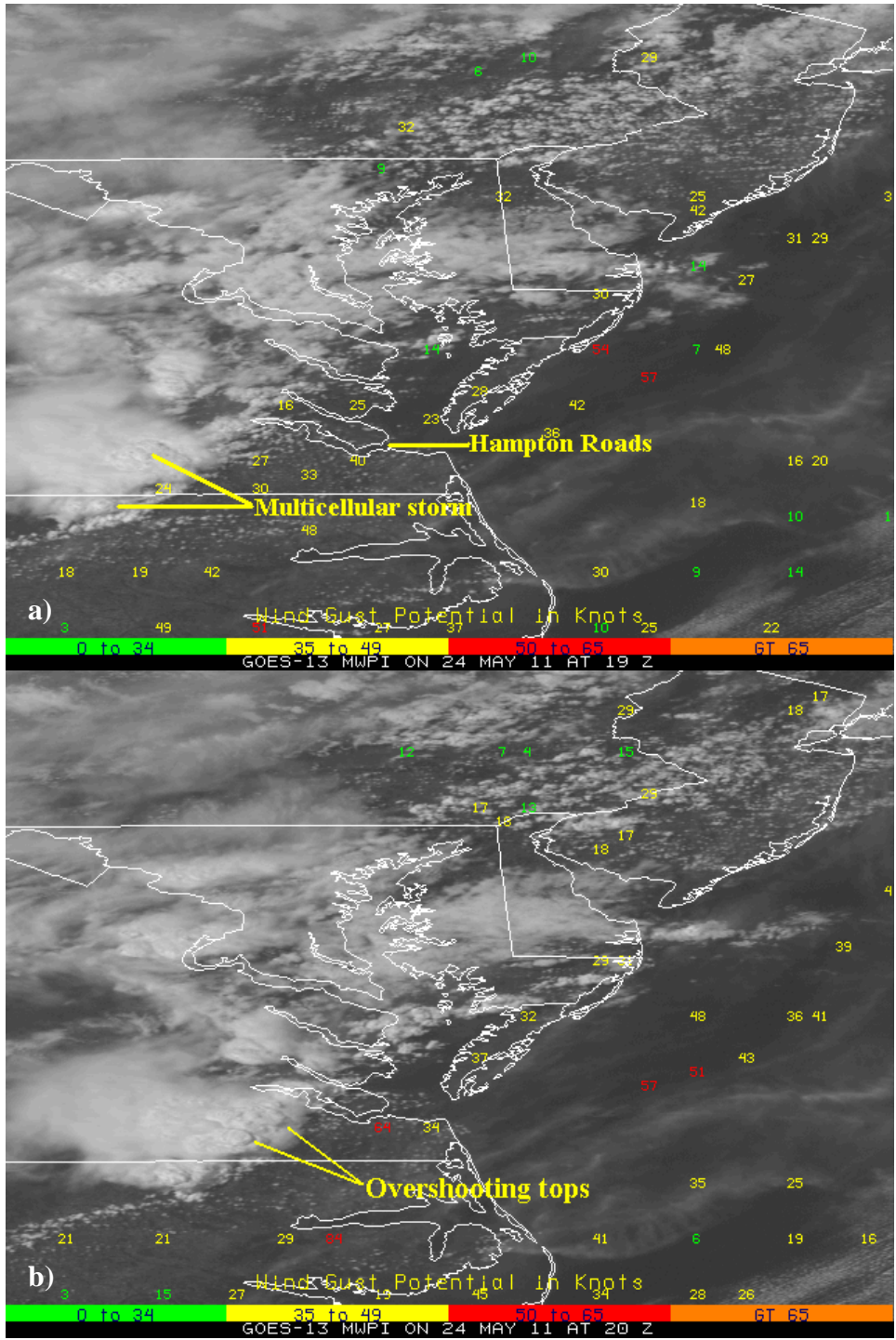

**Figure 4**. GOES MWPI products, with index values overlying visible imagery, at a) 1900 UTC and b) 2000 UTC 24 May 2011. A general increase in index values and associated downburst wind gust potential over Hampton Roads is apparent while the multicellular convective storm intensified on its track toward the lower Chesapeake Bay.

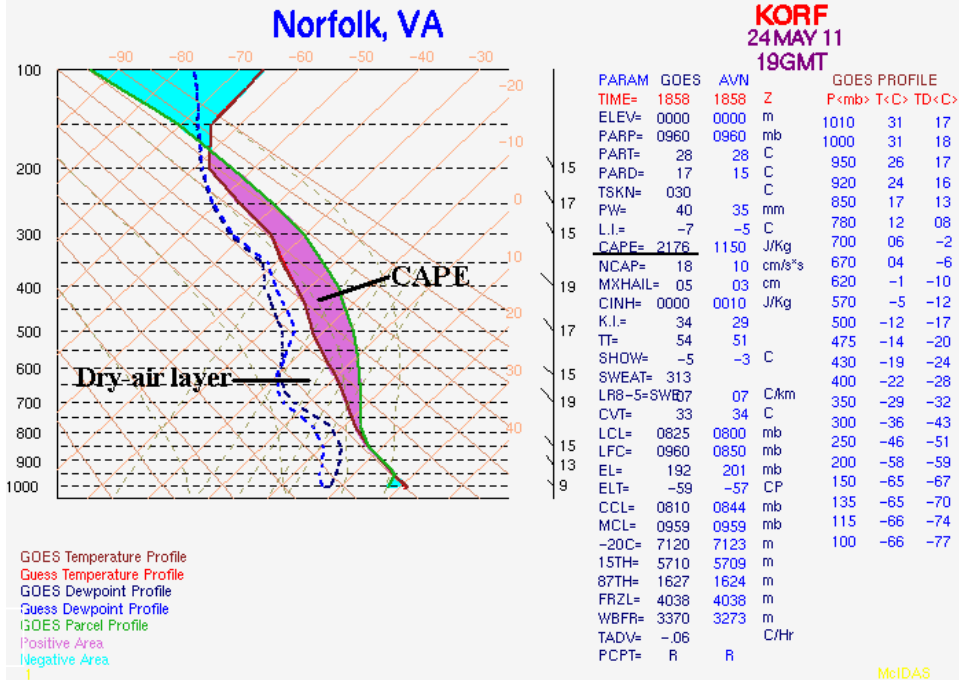

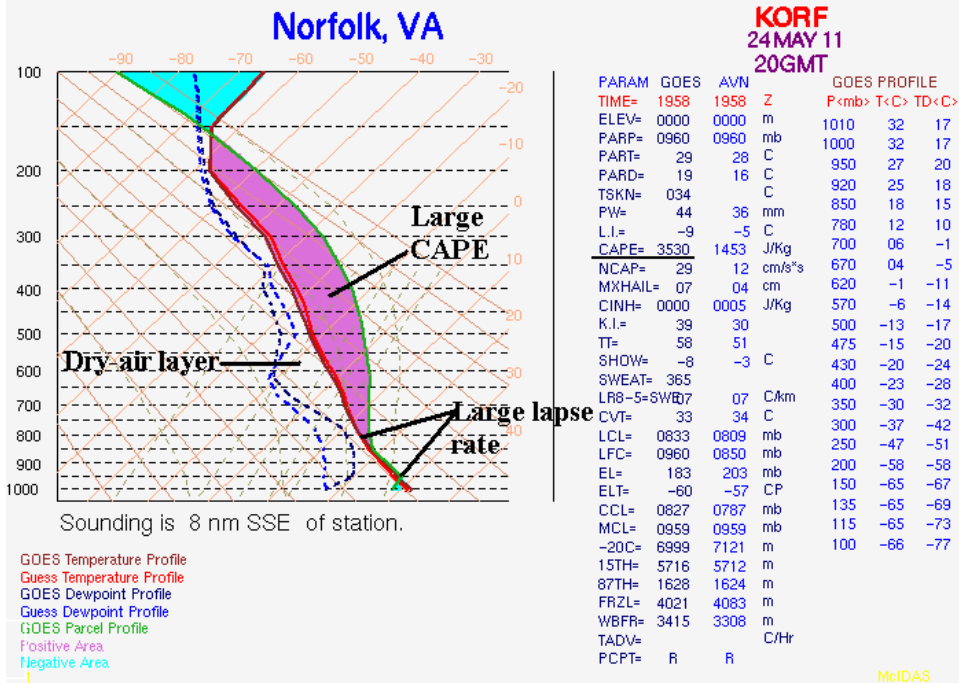

**Figure 5**. GOES sounding profiles over Norfolk, Virginia at a) 1900 UTC and b) 2000 UTC 24 May 2011.

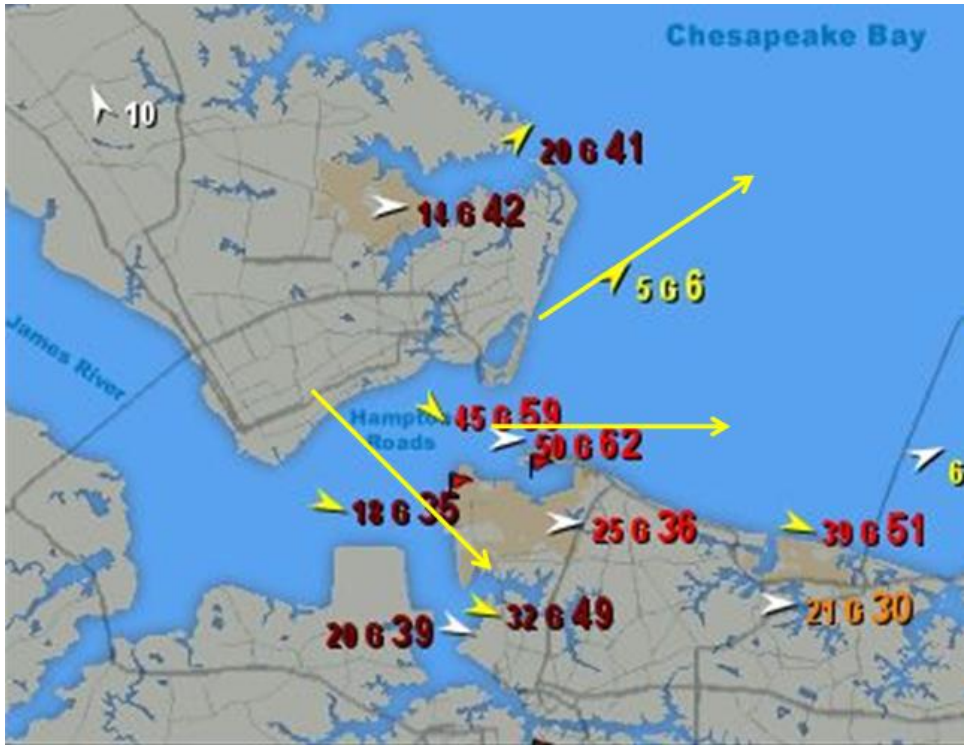

**Figure 6**.  WeatherFlow surface observation plot at 2037 UTC 24 May 2011 showing the divergent nature of convective storm outflow winds over Hampton Roads area.  Wind speeds are in knots.  Courtesy Weatherflow Datascope.

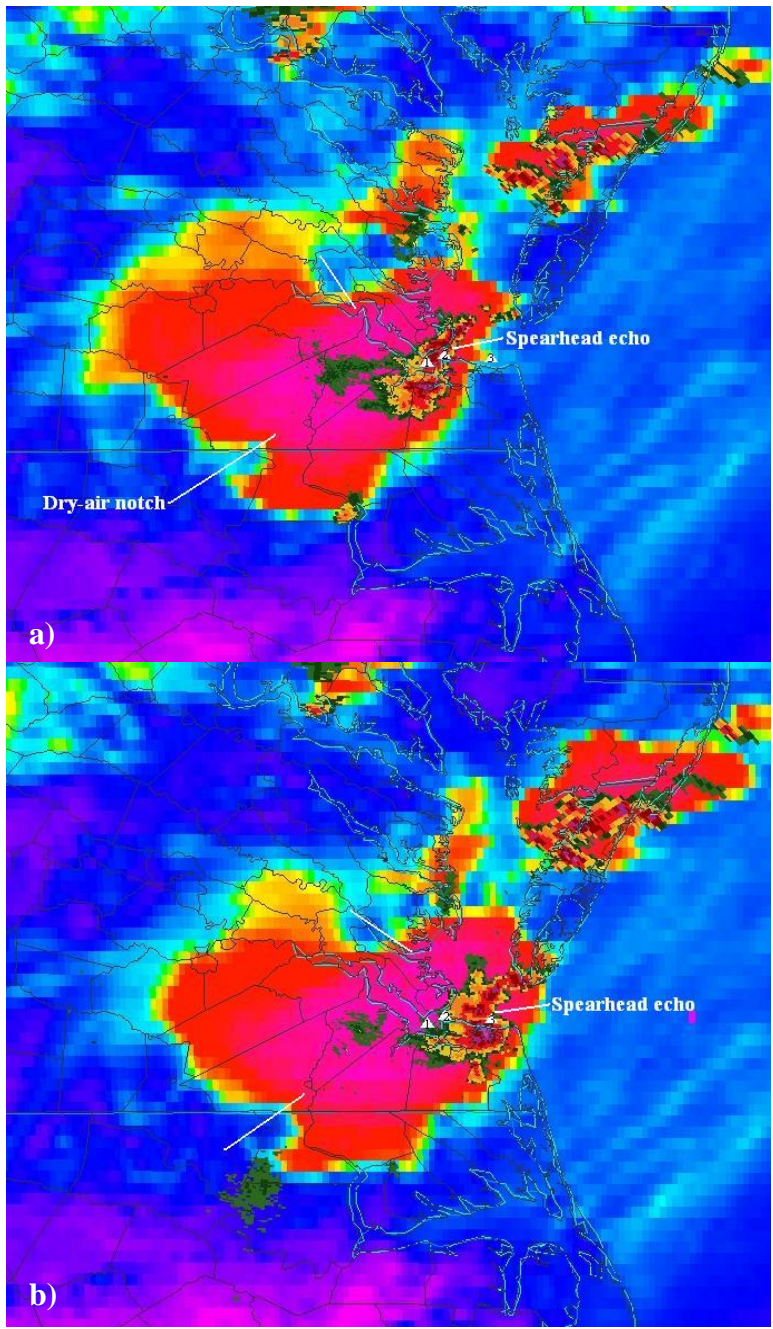

**Figure 7**. GOES imager channel 3-4 BTD products at a) 2025 UTC and b) 2040 UTC 24 May 2011 with overlying radar reflectivity from Wakefield, Virginia NEXRAD. Triangular markers indicate the location of 1) Monitor-Merrimac Memorial Bridge-Tunnel, 2) Willoughby Degaussing Station, and 3) Chesapeake Bay Bridge-Tunnel observing stations, respectively. White lines indicate the presence of dry-air notches.

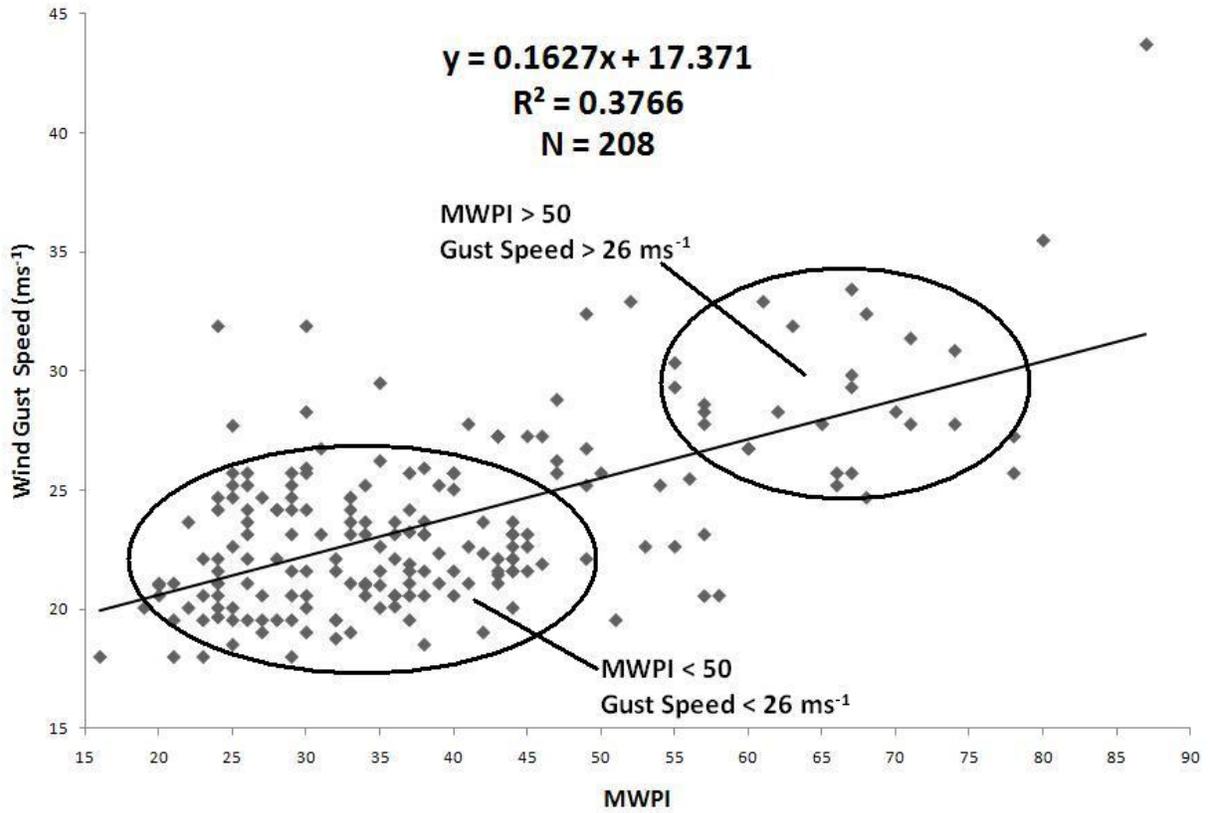

**Figure 8**. Statistical analysis of validation data over the Oklahoma and western Texas domain between June and September 2007 through 2010: Scatterplot of MWPI values vs. measured convective wind gusts for 208 downburst events.

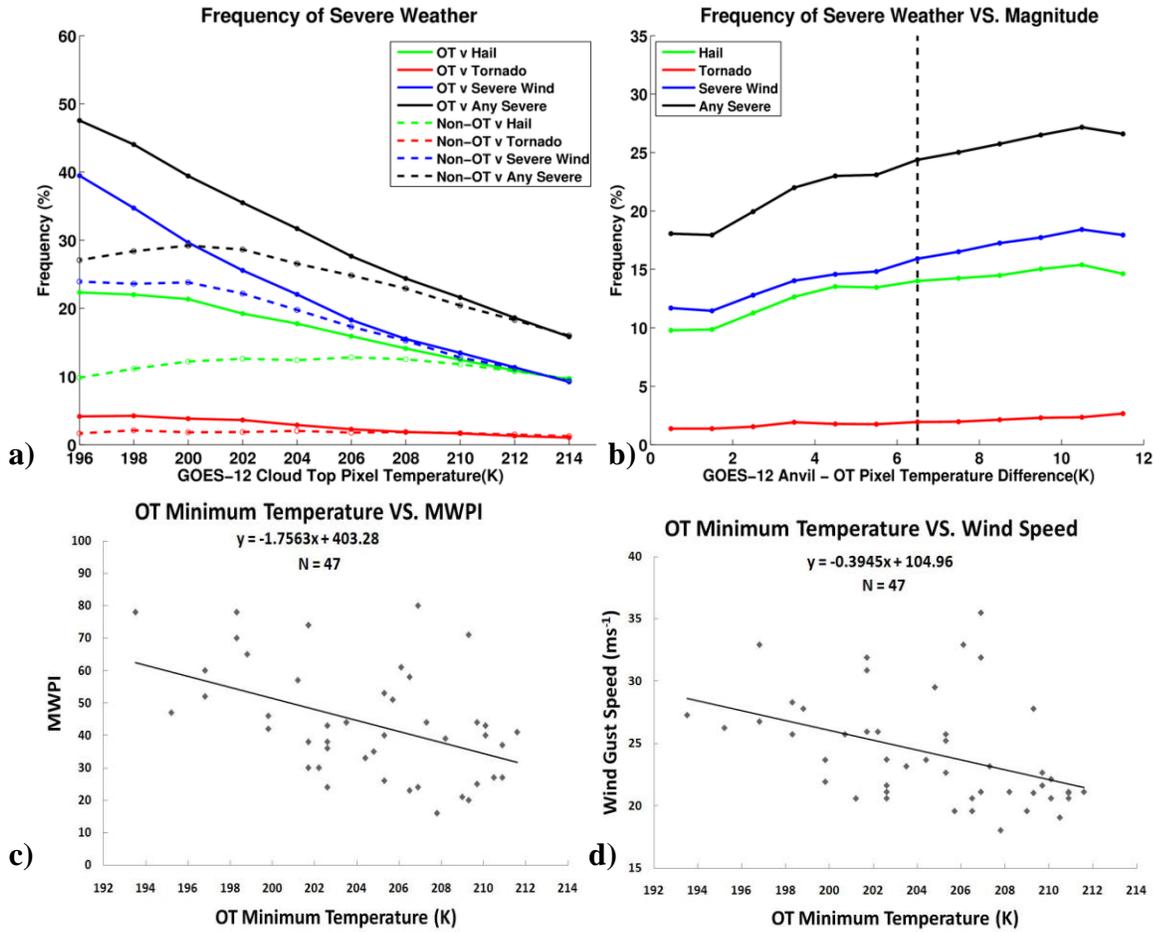

**Figure 9**. a) The frequency of severe weather for OTs (solid lines) and non-OT cold pixels (dashed lines) with varying IRW BT for each of the severe weather categories during the 2004-2009 warm seasons. b) Frequency of severe weather with varying BT difference between a pixel and the mean surrounding anvil temperature for each of the severe weather categories. The dashed line delineates the 6.5 °K criteria required for a pixel to be considered an OT. c) Comparison of scatterplots of convective OT minimum BT vs. GOES MWPI values and d) OT minimum BT vs. measured downburst wind gust speed (bottom) for 47 cases that occurred between 2007 and 2009.